Freedom in the Expansion and Obstacles to Integrability in Multiple-Soliton Solutions of the Perturbed KdV Equation

Alex Veksler[1] and Yair Zarmi[1,2]
Ben-Gurion University of the Negev, Israel
[1]Department of Physics, Beer-Sheva, 84105
[2]Department of Energy & Environmental Physics
Jacob Blaustein Institute for Desert Research, Sede-Boqer Campus, 84990

ABSTRACT

The construction of a solution of the perturbed KdV equation encounters *obstacles to asymptotic integrability* beyond the first order, when the zero-order approximation is a multiple-soliton wave. In the standard analysis, the obstacles lead to the loss of integrability of the Normal Form, resulting in a zero-order term, which does not have the simple structure of the solution of the unperturbed equation. Exploiting the freedom in the perturbative expansion, we propose an algorithm that shifts the effect of the obstacles from the Normal Form to the higher-order terms. The Normal Form then remains integrable, and the zero-order approximation retains the multiple-soliton structure of the unperturbed solution. The obstacles are expressed in terms of symmetries of the unperturbed equation, and decay exponentially away from the soliton-interaction region. As a result, they generate a bounded correction term, which decays exponentially away from the origin in the *x*-*t* plane. The computation is performed through second order.

## I. Introduction

The KdV equation [1], to which a small perturbation is added, is often analyzed by the method of Normal Forms (NF) [2-8]. The motivation is the expectation that like the KdV equation, the NF, which is the evolution equation of the zero-order approximation to the solution of the perturbed equation, will be integrable and preserve the wave nature of the solution of the unperturbed equation. When the zero-order approximation is a single-soliton state, this expectation is borne out [4, 8]; the NF then merely updates the dispersion relation obeyed by the wave velocity [9-11]. The situation is different when one seeks a solution for which the zero-order term is a multiple-soliton state. Except for specific forms of the perturbation, from second order and onwards, the standard formalism generates terms, which cannot be accounted for by the perturbative expansion of the solution (the Near Identity Transformation – NIT) [3-8]. One is forced to include them in the NF [4, 7, 8], rendering the latter non-integrable (hence the term "obstacles to integrability"), and spoiling the simplicity of the zero-order solution.

The loss of integrability of the NF is a consequence of an assumption made in the standard NF expansion, that, starting from the first order, the terms in the NIT are differential polynomials in the zero-order approximation[1]. In this paper, we propose an alternative algorithm, which allows for an additional $t$- and $x$-dependence in the higher-order terms, a dependence that cannot be accounted for by differential polynomials in the zero-order approximation. The added freedom enables one to shift the effect of the obstacles to integrability from the NF to the NIT. The NF then remains integrable. Its solution (the zero-order approximation) retains the multiple-soliton character of the solution of the unperturbed equation. Moreover, just as in the single-soliton case, the NF merely updates the dispersion relation obeyed by the velocity of each soliton. Our algorithm yields obstacles in a "canonical" form, expressed in terms of symmetries of the KdV equation. These obstacles

[1] The problem arises already in the NF analysis of perturbed ODE's, where, customarily, higher-order corrections in the NIT are assumed to be functions of the zero-order term. This assumption works for systems of *autonomous* equations with a *linear unperturbed part*. Inconsistencies may emerge in all other cases, and are resolved by allowing for explicit time dependence in the NIT. The issue is discussed in part in [25]

vanish identically when computed for a single-soliton. (Obstacles that are not in the canonical form do not vanish in the single-soliton case. The fact that they do not emerge in that case is then discovered through explicit calculation.) In the multiple-soliton case, the canonical obstacles do not vanish identically. However, away from the soliton-interaction region (a finite domain around the origin in the *t*-*x* plane) they decay exponentially. As a result, they generate in the NIT a bounded, exponentially decaying tail. The analysis is carried out through second order.

**2. NF analysis of the perturbed KdV equation**

In this Section, we review general aspects of the NF analysis of the perturbed KdV equation[2]:

$$w_t = 6w\,w_x + w_{xxx} + e\left(30\,a_1\,w^2\,w_x + 10\,a_2\,w\,w_{xxx} + 20\,a_3\,w_x\,w_{xx} + a_4\,w_{5x}\right)$$
$$+e^2\begin{pmatrix} 140\,b_1\,w^3\,w_x + 70\,b_2\,w^2\,w_{xxx} + 280\,b_3\,w\,w_x\,w_{xx} \\ +14\,b_4\,w\,w_{5x} + 70\,b_5\,{w_x}^3 + 42\,b_6\,w_x\,w_{4x} + 70\,b_7 w_{xx}\,w_{xxx} + b_8\,w_{7x} \end{pmatrix} + O\left(e^3\right) \quad . \qquad (2.1)$$
$$\left(|e| \ll 1\right)$$

The unperturbed equation,

$$u_t = 6u\,u_x + u_{xx} \equiv S_2\left[u\right], \qquad (2.2)$$

is integrable [12-24]. Its pure soliton eigen-solutions are the single soliton,

$$u\left(t,x\right) = 2\,k^2 \big/ \cosh^2\left(k\left\{x + v_0\,t + x_0\right\}\right) \quad , \qquad (2.3)$$

where

$$v_0 = 4\,k^2 , \qquad (2.4)$$

as well as multiples, which may be expressed by the Hirota formula [12]. For example, the two-soliton solution is given by

$$u(t,x) = 2\partial_x^2 \ln\left\{1 + g_1(t,x) + g_2(t,x) + \left(\frac{k_1 - k_2}{k_1 + k_2}\right)^2 g_1(t,x) g_2(t,x)\right\}$$
$$\left(g_i(t,x) = \exp\left[2k_i\left(x + v_{0,i}\, t + x_{0,i}\right)\right]\right) \quad , \tag{2.5}$$

where each $v_{0,i}$ is related to $k_i$ by Eq. (2.4). If the initial conditions employed do not correspond to the pure soliton eigen-solutions, then the solution develops dispersive tails [22-24]. This paper focuses on the pure soliton case.

---

[2] The numerical coefficients in Eq. (2.1) have been chosen so as to conform to the structure of the symmetries of the unperturbed equation, discussed in the following.

Away from the interaction region of the solitons, Eq. (2.5) is asymptotically reduced to a sum of two single-soliton solutions (see Fig. 1):

$$u(t,x) \to 2k_1^{\,2} \big/ \cosh^2\left(k_1\left\{x + v_{0,1}\, t + x_{0,1}\right\}\right) + 2k_2^{\,2} \big/ \cosh^2\left(k_2\left\{x + v_{0,2}\, t + x_{0,2}\right\}\right) \ . \tag{2.6}$$

The deviation of the exact solution from the asymptotic form falls off exponentially as the distance from the interaction region grows.

Returning to Eq. (2.1), we assume a *Near Identity Transformation* (NIT) for $w$:

$$w = u + e\, u^{(1)} + e^2 u^{(2)} + O\left(e^{\,3}\right) \ . \tag{2.7}$$

The evolution of the zero-order term, $u(t,x)$, is governed by the *Normal Form* (NF), which is constructed form $S_n$, the *symmetries* of the unperturbed equation [2-8, 13-24]:

$$u_t = S_2[u] + e\, a_4\, S_3[u] + e^2\, b_8\, S_4[u] + O\left(e^{\,3}\right) \ . \tag{2.8}$$

$S_n$ obey the recursion relation [15-21]

$$S_{n+1}[u] = \partial_x^{\,2} S_n[u] + 4u S_n[u] + 2u_1 G_n[u] \ , \tag{2.9}$$

with

$$G_n[u] = \partial_x^{-1} S_n[u] \quad . \tag{2.10}$$

For the present analysis, we shall need

$$\begin{aligned} S_1 &= u_x \\ S_2 &= 6uu_x + u_{xxx} \\ S_3 &= 30u^2 u_x + 10uu_{xxx} + 20u_x u_{xx} + u_{5x} \\ S_4 &= 140u^3 u_x + 70uu_{xxx} + 280uu_x u_{xx} + 14uu_{5x} + 70u_x^3 + 42u_x u_{4x} + 70u_{xx} u_{xxx} + u_{7X} \end{aligned} \quad . \tag{2.11}$$

Eq. (2.8) is integrable; the single- and multiple-soliton solutions of Eq. (2.2), are also solutions of Eq. (2.8), with each velocity updated according to a dispersion relation [9-11]:

$$v = 4k^2 + 16e\,a_4 k^4 + 64e^2 b_8 k^6 + O\left(e^3\right) \quad . \tag{2.12}$$

Using Eqs. (2.7) and (2.8) in Eq. (2.1), one obtains in $O(e^n)$ the $n$'th-order *homological equation*, which determines $u^{(n)}$, the $n$'th-order term in the NIT. The first-order equation is:

$$u_t^{(1)}[u;t,x] - 6\partial_x\left(uu^{(1)}\right) - \partial_x{}^3 u^{(1)} + a_4 S_3[u] = 30a_1 u^2 u_x + 10a_2 uu_{xxx} + 20a_3 u_x u_{xx} + a_4 u_{5x} \quad . \tag{2.13}$$

The second-order homological equation has the following structure

$$\begin{aligned} u_t^{(2)}[u;t,x] - 6\partial_x\left(uu^{(2)}\right) - \partial_x{}^3 u^{(2)} + b_8 S_4[u] = \\ + 140b_1 u^3 u_x + 70b_2 u^2 u_{xxx} + 280b_3 uu_x u_{xx} \\ + 14b_4 uu_{5x} + 70b_5 u_x{}^3 + 42b_6 u_x u_{4x} + 70b_7 u_{xx} u_{xxx} + b_8 u_{7x} + Z^{(2)} \end{aligned} \quad . \tag{2.14}$$

In Eq.(2.14), $Z^{(2)}$ is the contribution of the first-order part of the solution, $u^{(1)}$, to the second-order homological equation. It is given in the Appendix (Eq. (A.1)). In Eqs. (2.13) and (2.14) and in the following Sections, partial derivatives with respect to $t$ and $x$ are applied to the dependence of $u^{(n)}$ ($n$ = 1, 2) on these variables both through $u(t,x)$, as well as through the explicit $t$-and $x$- dependence.

## 3. Review of standard NF analysis [2-8]

In the standard analysis, the higher-order terms in Eq. (2.7) are assumed to be differential polynomials in $u$, with no explicit dependence on $t$ and $x$. No obstacle is encountered in the first-order equation, Eq. (2.13), [2-8]. The structure of $u^{(1)}$ is

$$u^{(1)} = a u^2 + b q^{(1)} u_x + c u_{xx} \qquad \left( q^{(1)} \equiv \int_{-\infty}^{x} u(x,t) dx \right) \quad . \tag{3.1}$$

(All terms in $u^{(1)}$, must have a total *weight* of 4, when the weights assigned to $u$, $?_t$ and $?_x$, are 2, 3 and 1, respectively [13-19].) Eq. (2.13) is solved for $u^{(1)}$, yielding:

$$a = -5 a_1 + \tfrac{5}{3} a_3 + \tfrac{10}{3} a_4 \quad , \quad b = -\tfrac{10}{3} a_3 + \tfrac{10}{3} a_4 \quad , \quad c = -\tfrac{5}{2} a_1 + \tfrac{5}{3} a_3 + \tfrac{5}{6} a_4 \quad . \tag{3.2}$$

In second-order, assuming that $u^{(2)}$ is a differential polynomial in $u$, with no explicit dependence on $t$ and $x$, it must have weight 6. The formalism allows for the following differential polynomial:

$$\begin{aligned} u^{(2)} = {} & A u^3 + B u^2 \left(q^{(1)}\right)^2 + C u u_x q^{(1)} + D u u_{xx} + E u \left(q^{(1)}\right)^4 + F u q^{(1)} q^{(2)} \\ & + G u_x^{\,2} + H u_x \left(q^{(1)}\right)^3 + I u_x q^{(2)} + J u_{xx} \left(q^{(1)}\right)^2 + K u_{xxx} q^{(1)} + L u_{xxxx} \quad . \\ & \left(q^{(1)} = \partial_x^{-1} u\right) \qquad \left(q^{(2)} = \partial_x^{-1} (u)^2\right) \end{aligned} \tag{3.3}$$

Despite the wealth of coefficients in Eq. (3.3), not all the terms on the r.h.s. of Eq. (2.14) can be accounted for. The unaccounted-for terms constitute the obstacle, $R^{(2)}$. The structure of the latter depends on the choice of coefficients. For example, if one chooses to account for as many terms on the r.h.s. of Eq. (2.14) as possible, the coefficients in Eq. (3.3) obtain the values given in the Appendix (Eqs. (A.2-10)), and the unaccounted-for term is:

$$R_{St}^{(2)} = m u^3 u_x \quad . \tag{3.4}$$

The coefficient $m$ is given in Eq. (A.11). The subscript *St* stands for the *standard analysis*. The structure of this term (with $m$ omitted) is shown in Fig. 2 for the two-soliton case. In general, the structure of other forms of the obstacle is similar to that of $R_{St}^{(2)}$.

As the obstacle, $R^{(2)}$ cannot be accounted for by the NIT in Eq. (2.14), one is forced to include it in the NF. Thus, against one's expectation, Eq. (2.8) has to be modified into

$$u_t = S_2[u] + e\,a_4 S_3[u] + e^2\left\{b_8 S_4[u] + R^{(2)}\right\} \ . \tag{3.5}$$

Obstacles spoil the integrability of the NF, because they cannot be written as linear combinations (with constant coefficients) of symmetries. Whereas Eq. (2.8) is solved by the same single- or multiple-soliton solutions of the unperturbed equation (with updated velocities), Eq. (3.5) is not. Its solution loses the KdV multiple-soliton structure, and may have to be computed numerically. For example, the obstacle in the analysis of [4, 8] for a two-soliton case leads to a zero-order solution that contains an $O(e^2)$ radiation term, an $O(e^4)$ time-dependent update of the wave numbers, and a new soliton with an $O(e^4)$ time-dependent wave number and an $O(e^8)$ amplitude.

## 4. Shifting the second-order obstacle from the NF to the NIT

The effect of the obstacle can be shifted from the NF to the NIT if one includes in $u^{(2)}$ an additional term, $u_r^{(2)}(t,x)$, which depends explicitly on $t$ and $x$:

$$u^{(2)} = u_d^{(2)}[u] + u_r^{(2)}(t,x) \ . \tag{4.1}$$

$u_d^{(2)}$ is the differential polynomial of Eq. (3.3). (Eqs. (A.2-10) provide an example of a possible choice of the coefficients in Eq. (3.3)). $u_r^{(2)}$ accounts for the second-order obstacle, $R^{(2)}$, through Eq. (2.14), which is reduced to an equation for $u_r^{(2))}$:

$$\partial_t u_r^{(2)}(t,x) = 6\partial_x\left\{u\, u_r^{(2)}(t,x)\right\} + \partial_x^3 u_r^{(2)}(t,x) + R^{(2)} \quad . \tag{4.2}$$

With the NF relieved of the burden of accounting for the obstacle, it retains its preferred form of Eq. (2.8) and remains integrable. The penalty paid for the loss of integrability is that, in general, the solution of Eq. (4.2) may not be expressible as a differential polynomial in $u$, and may have to be solved numerically.

The obstacle of Eq. (3.4), plotted in Fig. 2, may cause two problems, which have not been addressed yet. First, obstacles may not vanish when $u$ is a single-soliton solution, while there ought to be no obstacles in that case [4, 8]. (The reason is that some of the computational steps leading, for example, to Eq. (3.4) are not possible in the case of the single-soliton zero-order solution.) Second, in the multiple-soliton case, obstacles may overlap with the solution over an infinite range in $t$ and $x$, potentially leading to the generation of unbounded terms in the solution of Eq. (4.2). These two difficulties are resolved if $u_d^{(2)}$, the differential polynomial part in Eq. (4.1), is chosen to have the structure of $u_s^{(2)}$, the differential polynomial that solves the second-order homological equation in the case of a *single-soliton* solution. However, now it is computed for $u$, the solution of the NF in the general case. Eq. (4.1) then becomes:

$$u^{(2)} = u_s^{(2)}[u] + u_r^{(2)}(t,x) \quad . \tag{4.3}$$

With $u^{(2)}$ of Eq. (4.3), $u_s^{(2)}$ accounts for all the differential monomials in Eq. (2.14), that can be cancelled when $u$ is a single-soliton zero-order solution of the NF. As a result, the structure of the obstacle is such that, if computed for the single-soliton case, it vanishes explicitly. Moreover, in the multiple-soliton case, the resulting obstacle is expected to be sizable only in the interaction region, a finite domain around the origin, and to decay exponentially to zero away from the interaction region. The reason is that, away from the soliton interaction region, the zero-order solution ap-proaches asymptotically a sum of distinct single solitons at an exponential rate. These statements are demonstrated in Section 7, in the two-soliton case.

## 5. NF analysis in single-soliton case

For later use, we review the expansion of the solution of Eq. (2.1) for the case of the single-soliton solution of the NF, Eq. (2.8), given by Eq. (2.3). The only modification relative to the unperturbed single-soliton solution is that the velocity is updated according to Eq. (2.12). Only some of the coefficients in $u^{(1)}$ and in $u^{(2)}$ of Eq. (2.7) are determined, because differential monomials, which are considered independent in the general analysis, become related. (See Eqs. (A.12-13) for examples.)

In first order, one expects there to be no obstacle in the single-soliton case, because, already in the general case, a differential polynomial solution for $u^{(1)}$ exists, which accounts for all the terms in Eq. (2.13) and no obstacle emerges, provided *a*, *b* and *c* take on the values given in Eq. (3.2). In the single-soliton case, inserting Eqs. (2.8) and (3.1) in Eq. (2.13), one of the coefficients, *a*, *b* or *c*, in Eq. (3.1) remains undetermined. Choosing *c* as the free parameter, $u^{(1)}$ obtains the form

$$\begin{aligned} u_s^{(1)}[u] = & \frac{1}{6}\left(6c - 15\alpha_1 + 10\alpha_2 - 10\alpha_3 + 15\alpha_4\right)u^2 \\ & + \frac{1}{6}\left(6c + 15\alpha_1 - 20\alpha_2 - 10\alpha_3 + 15\alpha_4\right)q^{(1)}u_x + c\,u_{xx} \end{aligned} \quad . \qquad (5.1)$$

(The subscript *s* indicates that this is the case of a *single-soliton* zero-order solution.)

The solution of the second-order homological equation, Eq. (2.14), is also a differential polynomial, $u_s^{(2)}[u]$, and no obstacles to integrability emerge [4, 8]. The values of those coefficients in Eq. (3.3) which can be determined are given in the Appendix (Eqs. (A.14-16)); the remaining ones are free.

**6. Absence of first-order obstacle in the general case and “canonical” obstacles**

In Section 3, we reviewed the known observation that no obstacles emerge in the first-order homological equation, Eq. (2.13), even in the general case. It is instructive to see how this comes about. In the analysis of Eq. (2.13) in the general case, the coefficients $a$, $b$ and $c$ of the expression for $u^{(1)}$ (see Eq. (3.1)) receive the values given in Eq. (3.2). In the single-soliton case, $u^{(1)}$ assumes the form given by Eq. (5.1), and there is no first-order obstacle, although one parameter, e.g., $c$, remains free. Let us, for the moment, adopt Eq. (5.1) for $u^{(1)}$ also for the general case. Before $c$ is assigned the value specified by Eq. (3.2), some differential monomials in Eq. (2.13) are not accounted for; an “obstacle” emerges, given by:

$$R^{(1)} = g_{21}^{(1)} R_{21} \qquad \left( g_{21}^{(1)} = \frac{1}{2}\left(6c + 15a_1 - 10a_3 - 5a_4\right) \right) \quad . \tag{6.1}$$

In Eq. (6.1), $R_{21}$ has the form

$$R_{21} = 3u^2 u_x + u u_{xxx} - u_x u_{xx} = S_2 G_1 - G_2 S_1 \quad . \tag{6.2}$$

(The notation for the subscript used for the obstacle is self-evident.) Not surprisingly, with $c$ given in Eq. (3.2), this first-order obstacle is eliminated. One also sees why $c$ remains free in the single-soliton case: As shown in the following, $R_{21}$ vanishes if computed for the single-soliton solution of Eq. (2.3).

$R_{21}$ is the first example of the “canonical” obstacles, defined as

$$R_{nm} = S_n G_m - G_n S_m \quad . \tag{6.3}$$

Obstacles encountered in the higher-order analysis of the multiple-soliton case will be linear functionals of $R_{\mathrm{nm}}$.

The importance of the canonical obstacles is that all $R_{nm}$ vanish when computed for the *single-soliton* solution. This property is a consequence of the recursion relation, Eq. (2.9), obeyed by the symmetries, $S_n$. To see this, consider, first, the single-soliton solution of the unperturbed KdV equation, which we write as

$$u(t,x) = u(x_0) \qquad x_0 = x + v_0 t \quad \left(v_0 = 4k^2\right) \ . \tag{6.4}$$

Consequently, the unperturbed equation, Eq. (2.2), can be re-written as

$$v_0 \, \partial_{x_0} u = S_2[u] \ . \tag{6.5}$$

Eq. (6.5) is a relation between the first two symmetries in the KdV hierarchy (see Eq. (2.11)):

$$v_0 \, S_1[u] = S_2[u] \ . \tag{6.6}$$

For localized soliton solutions that vanish at $x_0 = \pm 8$, $G_n$, defined in Eq. (2.10), obey a similar relation:

$$G_2[u] = 3u^2 + u_{xx} = v_0 \, G_1[u] = v_0 u \ . \tag{6.7}$$

Substituting Eqs. (6.6) and (6.7) in Eq. (2.9), one readily obtains by induction

$$S_n[u] = (v_0)^{n-1} S_1[u] \ , \qquad G_n[u] = (v_0)^{n-1} G_1[u] \ . \tag{6.8}$$

The single-soliton zero-order solution of Eq. (2.8) has the *same* functional form as the unperturbed solution, Eq. (2.3), the only change being that the velocity is modified by the dispersion relation, Eq. (2.12). As the derivation of Eq. (6.8) involves only $u$ and its spatial derivatives, it remains valid for the single-soliton solution in the perturbed case as well.

The proportionality of all symmetries to $S_1$ implies that the canonical obstacles vanish explicitly in the single-soliton case. More important, in the multiple-soliton case, the solution approaches a sum of well-separated single solitons at an exponential rate as the distance from the soliton-interaction region grows. Consequently, the canonical obstacles vanish asymptotically at the same rate.

## 7. NF analysis in two-soliton case

We now present the NF analysis of Eq. (2.1) when the zero-order approximation (solution of Eq. (2.8)) is the two-soliton solution, Eq. (2.5), in which the velocities are updated through Eq. (2.12). We use $u_s^{(1)}$ of Eq. (3.1), with the coefficients given by Eq. (3.2), and $u^{(2)}$ of Eq. (4.3). The coefficients of $u_s^{(2)}$ are given in Eqs. (A.14-A.16). A number of them remain free. The second-order homological equation, Eq. (2.14), determines some of these coefficients. The resulting coefficients are given in the Eq. (A.17-A.19). The structure of the second-order obstacle is found to be

$$R^{(2)} = \left\{ Q_1\, u + Q_2\, \partial_x^2 + Q_3\, u_x\, \partial_x^{-1} \right\} R_{21} + Q_4\, R_{31} \quad . \tag{7.1}$$

The coefficients $Q_i$, $1 = i = 4$, of Eq. (7.1) are given in Eqs. (A.20-23). As $R^{(2)}$ is constructed from canonical obstacles, it vanishes identically in the single-soliton case. In the multiple-soliton case, it decays exponentially away from the interaction region of the solitons. An appropriate choice of the still free coefficients in Eq. (3.3) exists (given in Eqs. (A.24-26)), which simplifies Eq. (7.1) into:

$$R^{(2)} = -\tfrac{10}{3}\, m\, u\, R_{21} \quad . \tag{7.2}$$

$m$ is given in Eq. (A.11). The obstacle $R^{(2)}$ is accounted for by $u_r^{(2)}(t,x)$, through Eq. (4.2)

For $|x|$, $|t|\to\infty$, the obstacle vanishes exponentially. (Fig. 3 demonstrates this property.) Substituting the expression for the two-soliton solution, Eq. (2.5), in Eq. (7.2), one finds after a detailed inspection that the asymptotic behavior of the obstacle along soliton no. 1 is:

$$R_{21} \propto e^{-2\left|k_1\left(x+4k_1^2 t\right)\right|} \qquad \left|k_1\left(x+4k_1^2 t\right)\right| \to \infty, \quad \left|k_2\left(x+4k_2^2 t\right)\right| = C \quad , \tag{7.3}$$

and a similar behavior along soliton no. 2. Far from both solitons, the behavior is

$$R_{21} \propto \max\left\{ e^{-2\left|k_1\left(x+4k_1^2 t\right)\right| - 4\left|k_2\left(x+4k_2^2 t\right)\right|}, \quad e^{-4\left|k_1\left(x+4k_1^2 t\right)\right| - 2\left|k_2\left(x+4k_2^2 t\right)\right|} \right\} \quad . \tag{7.4}$$
$$\left|k_1\left(x+4k_1^2 t\right)\right| \to \infty, \quad \left|k_2\left(x+4k_2^2 t\right)\right| \to \infty$$

The dispersion relation obeyed by these asymptotic terms does not resonate with the homogeneous part of Eq. (4.2). Hence, they generate a bounded tail, which emanates from the soliton-interaction region around the origin, and decays exponentially as the distance from the origin grows.

We end this Section with a comment regarding higher orders in the expansion. If the $n$'th order term in the NIT is constructed according to the algorithm used in the second analysis, namely,

$$u^{(n)} = u_s^{(n)}[u] + u_r^{(n)}(t,x) \quad , \tag{7.5}$$

where $u_s^{(n)}$ is the differential polynomial obtained as the solution of the $n$'th-order homological equation in the *single-soliton* case (now computed for the two-soliton solution), then the obstacle is guaranteed to vanish explicitly in the single-soliton case. Eq. (7.1) suggests that the higher-order obstacles are expected to be linear functionals of canonical obstacles:

$$R^{(n)}[u] = \sum g_{pq}^{(n)} f_{pq}^{(n)}\left[u, \partial_x, \partial_x^{-1}\right] R_{pq}[u] \quad . \tag{7.6}$$

The coefficients $g_{pq}^{(n)}$ are combinations of the perturbation parameters. $u_r^{(n)}(t,x)$ of Eq. (7.3) accounts for the obstacle, so that the NF is unaffected by it. As the obstacle vanishes at an exponential rate away from the soliton interaction region, it cannot generate unbounded behavior in the approximate solution.

## 8. Concluding remarks

The effect of obstacles to integrability in the perturbed KdV equation can be shifted from the NF to the NIT by allowing the higher-order corrections in the NIT to depend explicitly on $t$ and $x$. The penalty for the loss of integrability is the fact that the NIT ceases to be a sum of differential polynomials in the zero-order approximation; some parts of the NIT may have to be found numerically. The gain is that the NF is constructed from symmetries only, hence, remains integrable, and the zero-order term retains the multiple-soliton structure of the unperturbed solution.

The "canonical" obstacles generated by our algorithm (see Eq. (7.5)) are expressed in terms of symmetries of the unperturbed equation and vanish explicitly in the case of a single-soliton solution of the NF. Moreover, as the canonical obstacles decay rapidly away from the interaction region in the multiple-soliton case, they do not generate unbounded behavior in the solution, but only a decaying tail, which emanates from the soliton-interaction region around the origin.

This conclusion is in agreement with the numerical studies of [26]. That work shows that the numerical solution of the ion acoustic plasma equations for a two-soliton collision is described extremely well by the two-soliton solution of the KdV equation, which is derived as an approximation to the ion acoustic plasma equations. The difference between the full equations and the KdV equation can be viewed as a small perturbation that is added to the latter. The result of [26] implies that the perturbation does not alter the nature of the solution through second order. A dispersive wave found by [26] seems to be consistent with a third-order effect.

**Appendix**

$Z^{(2)}$ of Eq. (2.14)

$$Z^{(2)} = 6u^{(1)}u^{(1)}{}_x + 30a_1\left(2uu_x u^{(1)} + u^2 u^{(1)}{}_x\right) + 10a_2\left(u_{xxx}u^{(1)} + uu^{(1)}{}_{xxx}\right) \\ + 20a_3\left(u_{xx}u^{(1)}{}_x + u_x u^{(1)}{}_{xx}\right) + a_4 u^{(1)}{}_{xxxxx} \quad . \qquad \text{(A.1)} \\ \left(u^{(1)} = u^{(1)}[u;t,x]\right)$$

Here derivatives of $u^{(1)}[u;t,x]$ with respect to $x$ account for its dependence on $x$ both through $u$ as well as through the explicit dependence on $x$.

Coefficients in Eq. (3.3): Standard NF analysis

$$A = \tfrac{50}{9}\left(9a_1^{\ 2} - 3a_1a_2 + a_2^{\ 2} - 6a_1a_3 + 20a_2a_3 - 8a_3^{\ 2} - 21a_2a_4 + 8a_4^{\ 2}\right) \\ - \tfrac{14}{3}\left(5b_2 + 30b_3 - 29b_4 - 20b_5 + 30b_6 - 20b_7 + 4b_8\right) \quad , \qquad \text{(A.2)}$$

$$C = \tfrac{100}{3}a_1\left(a_2 - a_4\right) - 28\left(b_4 - b_8\right) \quad , \qquad \text{(A.3)}$$

$$D = \tfrac{25}{9}\begin{pmatrix} 27a_1^{\ 2} - 3a_1a_2 + 4a_2^{\ 2} - 30a_1a_3 + 58a_2a_3 - 24a_3^{\ 2} \\ + 3a_1a_4 - 63a_2a_4 + 8a_3a_4 + 20a_4^{\ 2} \end{pmatrix} \\ - \tfrac{14}{3}\left(10b_2 + 40b_3 - 41b_4 - 30\,b_5 + 45\,b_6 - 30b_7 + 6b_8\right) \quad , \qquad \text{(A.4)}$$

$$G = \tfrac{25}{9}\begin{pmatrix} 18a_1^{\ 2} + 3a_1a_2 + 3a_2^{\ 2} - 24a_1a_3 + 40a_2a_3 - 16a_3^{\ 2} \\ + 3a_1a_4 - 48a_2a_4 + 6a_3a_4 + 15a_4^{\ 2} \end{pmatrix} \\ - \tfrac{7}{3}\left(15b_2 + 60b_3 - 59b_4 - 45b_5 + 60b_6 - 40b_7 + 9b_8\right) \quad , \qquad \text{(A.5)}$$

$$I = \tfrac{50}{9}\left(3a_1a_2 + a_2^{\ 2} - 7a_2a_4 + 3a_4^{\ 2}\right) - \tfrac{14}{3}\left(5b_2 - 7b_4 + 2b_8\right) \quad , \qquad \text{(A.6)}$$

$$J = \tfrac{50}{9}\left(a_2 - a_4\right)^2 \quad , \qquad \text{(A.7)}$$

$$K = \tfrac{25}{9}\left(a_2 - a_4\right)\left(3a_1 - 2a_3 + a_4\right) - \tfrac{14}{3}\left(b_4 - b_8\right) \quad , \qquad \text{(A.8)}$$

$$L = \tfrac{25}{72}\begin{pmatrix} 27a_1^{\ 2} + 4a_2^{\ 2} - 36a_1a_3 + 56a_2a_3 - 20a_3^{\ 2} + 6a_1a_4 \\ - 64a_2a_4 + 4a_3a_4 + 23a_4^{\ 2} \end{pmatrix} \\ - \tfrac{7}{6}\left(5b_2 + 20\,b_3 - 20b_4 - 15b_5 + 21b_6 - 15b_7 + 4b_8\right) \quad , \qquad \text{(A.9)}$$

$$B = E = F = H = 0 \quad . \qquad \text{(A.10)}$$

Coefficient *m* of obstacle in standard analysis (Eqs. (3.4) & (3.5))

$$\begin{aligned} m = \tfrac{100}{9}\left(3a_1a_2 + 4a_2^{\,2} - 18a_1a_3 + 60a_2a_3 - 24a_3^{\,2} + 18a_1a_4 - 67a_2a_4 + 24a_4^{\,2}\right) \\ + \tfrac{140}{3}\left(3b_1 - 4b_2 - 18b_3 + 17b_4 + 12b_5 - 18b_6 + 12b_7 - 4b_8\right) \end{aligned} \quad . \qquad \text{(A.11)}$$

Examples of relations obeyed by single-soliton solution

$$\partial_x\left(u\,q^{(1)}\right) = u^2 + u\left(q^{(1)}\right)^{(2)} \quad , \qquad \text{(A.12)}$$

$$u_{xx} = -2u^2 + \tfrac{1}{2}u\,q^{(1)} \quad . \qquad \text{(A.13)}$$

Coefficients in Eq. (3.3) in single-soliton case

$$\begin{aligned} A = \tfrac{5}{12}&\begin{pmatrix} 105a_1^{\,2} - 108a_1a_2 - 24a_2^{\,2} + 28a_1a_3 - 80a_2a_3 + 4a_3^{\,2} \\ -78a_1a_4 + 192a_2a_4 + 60a_3a_4 - 99a_4^{\,2} \end{pmatrix} \\ &- \tfrac{7}{3}\left(11b_1 - 18b_2 - 6b_3 + 18b_4 + 4b_5 - 9b_6 + 9b_7 - 9b_8\right) \\ &- B + C - E - \tfrac{4}{3}F - G + H + \tfrac{4}{3}I - 3K + 6L \end{aligned} \quad , \qquad \text{(A.14)}$$

$$\begin{aligned} D = \tfrac{5}{36}&\begin{pmatrix} 225a_1^{\,2} - 252a_1a_2 - 56a_2^{\,2} + 12a_1a_3 - 200a_2a_3 - 44a_3^{\,2} \\ -162a_1a_4 + 528a_2a_4 + 300a_3a_4 - 351a_4^{\,2} \end{pmatrix} \\ &- \tfrac{7}{3}\left(8b_1 - 14b_2 - 8b_3 + 17b_4 + 2b_5 - 3b_6 + 7b_7 - 9b_8\right) \\ &- B + C - E - \tfrac{4}{3}F - G + H + \tfrac{4}{3}I - 3K + 10L \end{aligned} \quad , \qquad \text{(A.15)}$$

$$\begin{aligned} J = \tfrac{5}{72}&\begin{pmatrix} 135a_1^{\,2} - 216a_1a_2 + 52a_2^{\,2} - 84a_1a_3 + 40a_2a_3 + 28a_3^{\,2} \\ +54a_1a_4 + 24a_2a_4 - 60a_3a_4 + 27a_4^{\,2} \end{pmatrix} \\ &- \tfrac{7}{6}\left(4b_1 - 7b_2 - 4b_3 + 8b_4 + b_5 - 3b_6 + b_7\right) \\ &- E - \tfrac{1}{3}F + H + \tfrac{1}{3}I + K - L \end{aligned} \quad . \qquad \text{(A.16)}$$

Constraints on coefficients in Eq. (3.3) in general case, with $u_S^{(2)}$ differential polynomial in Eq. (4.3)

$$B = E = F = H = 0 \quad , \qquad \text{(A.17)}$$

$$K = \tfrac{25}{9}\left(a_1a_2 - 2a_2a_3 - a_1a_4 + a_2a_4 + 2a_3a_4 - a_4^{\,2}\right) + \tfrac{1}{6}C \quad , \qquad \text{(A.18)}$$

$$\begin{aligned} L = \tfrac{5}{72}&\begin{pmatrix} 135a_1^{\,2} - 176a_1a_2 - 28a_2^{\,2} - 84a_1a_3 - 40a_2a_3 + 28a_3^{\,2} \\ +14a_1a_4 + 224a_2a_4 + 20a_3a_4 - 93a_4^{\,2} \end{pmatrix} \\ &+ \tfrac{1}{6}C + \tfrac{1}{3}I \end{aligned} \quad . \qquad \text{(A.19)}$$

Coefficients of obstacle in Eq. (7.1)

$$Q_1 = -\tfrac{5}{6}\begin{pmatrix} 630a_1^{\ 2} - 889a_1a_2 - 167a_2^{\ 2} - 336a_1a_3 - 280a_2a_3 + 112a_3^{\ 2} \\ +31a_1a_4 + 1316a_2a_4 + 210a_3a_4 - 627a_4^{\ 2} \end{pmatrix} + \tfrac{7}{2}\left(84b_1 - 167b_2 - 84b_3 + 201b_4 + 21b_5 - 84b_6 + 56b_7 - 27b_8\right) - \tfrac{43}{4}C + \tfrac{21}{2}G - 24I \quad , \tag{A.20}$$

$$Q_2 = -\tfrac{5}{6}\begin{pmatrix} 90a_1^{\ 2} - 127a_1a_2 - 21a_2^{\ 2} - 48a_1a_3 - 40a_2a_3 + 16a_3^{\ 2} \\ +13a_1a_4 + 168a_2a_4 + 30a_3a_4 - 81a_4^{\ 2} \end{pmatrix} + \tfrac{7}{2}\left(12b_1 - 21b_2 - 12b_3 + 23b_4 + 3b_5 - 12b_6 + 8b_7 - b_8\right) - \tfrac{7}{4}C + \tfrac{3}{2}G - 3I \quad , \tag{A.21}$$

$$Q_3 = -\tfrac{25}{6}\begin{pmatrix} 54a_1^{\ 2} - 77a_1a_2 - 15a_2^{\ 2} - 24a_1a_3 - 40a_2a_3 + 16a_3^{\ 2} \\ -a_1a_4 + 128a_2a_4 + 18a_3a_4 - 59a_4^{\ 2} \end{pmatrix} + \tfrac{35}{2}\left(8b_1 - 15b_2 - 12b_3 + 21b_4 + 5b_5 - 12b_6 + 8b_7 - 3b_8\right) - \tfrac{19}{4}C + \tfrac{9}{2}G - 10I \quad , \tag{A.22}$$

$$Q_4 = \tfrac{5}{6}\begin{pmatrix} 90a_1^{\ 2} - 107a_1a_2 - 21a_2^{\ 2} - 48a_1a_3 - 40a_2a_3 + 16a_3^{\ 2} \\ -7a_1a_4 + 168a_2a_4 + 30a_3a_4 - 81a_4^{\ 2} \end{pmatrix} - \tfrac{7}{2}\left(12b_1 - 21b_2 - 12b_3 + 27b_4 + 3b_5 - 12b_6 + 8b_7 - 5b_8\right) \tfrac{5}{4}C - \tfrac{3}{2}G + 3I \quad . \tag{A.23}$$

Choice of coefficients in Eq. (3.3) that eliminates $Q_2$, $Q_3$, $Q_4$ of Eq. (7.1)

$$C = \tfrac{100}{3}a_1\left(a_2 - a_4\right) - \tfrac{84}{3}\left(b_4 - b_8\right) \quad , \tag{A.24}$$

$$G = \tfrac{25}{9}\begin{pmatrix} 18a_1^{\ 2} + 3a_1a_2 + 3a_2^{\ 2} - 24a_1a_3 + 40a_2a_3 - 16a_3^{\ 2} \\ +3a_1a_4 - 48a_2a_4 + 6a_3a_4 + 15a_4^{\ 2} \end{pmatrix} - \tfrac{7}{3}\left(15b_2 + 60b_3 - 59b_4 - 45b_5 + 60b_6 - 40b_7 + 9b_8\right) \quad , \tag{A.25}$$

$$I = \tfrac{10}{3}\begin{pmatrix} 6a_1a_2 + 3a_2^{\ 2} - 6a_1a_3 + 20a_2a_3 - 8a_3^{\ 2} \\ +6a_1a_4 - 34a_2a_4 + 13a_4^{\ 2} \end{pmatrix} + 14\left(b_1 - 3b_2 - 6b_3 + 8b_4 + 4b_5 - 6b_6 + 4b_7 - 2b_8\right) \quad . \tag{A.26}$$

Figure captions

Fig. 1  Two-soliton solution (Eq. (2.5); $k_1 = 0.3$, $k_2 = 0.4$, $\delta_1 = \delta_2 = 0$.

Fig. 2  Obstacle for two-soliton solution in standard NF analysis (Eq. (3.4)); parameters as in Fig. 1.

Fig. 3  Canonical obstacle $u \cdot R_{21}$ (Eq.(6.2)) for two-soliton solution; parameters as in Fig. 1.

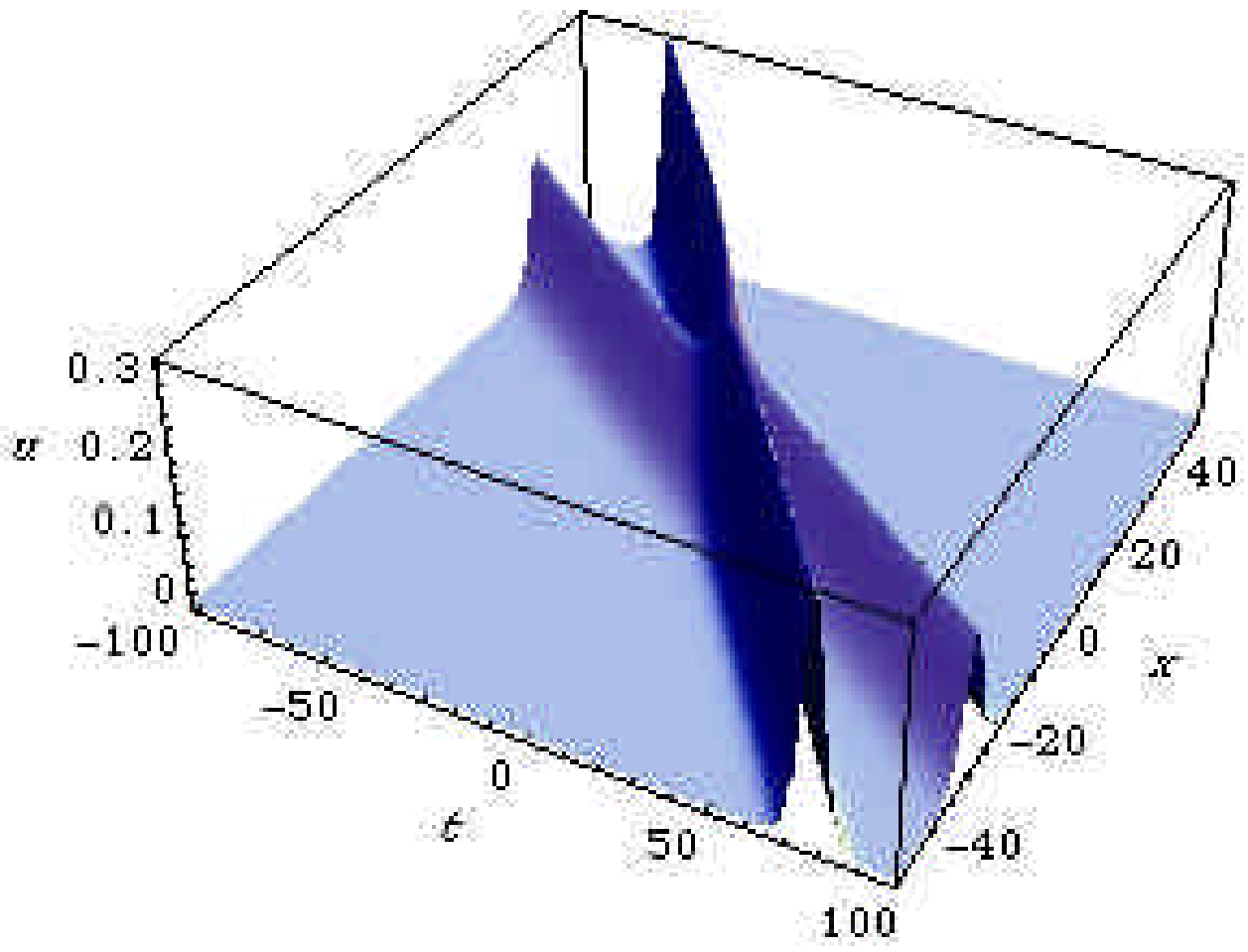

Fig. 1

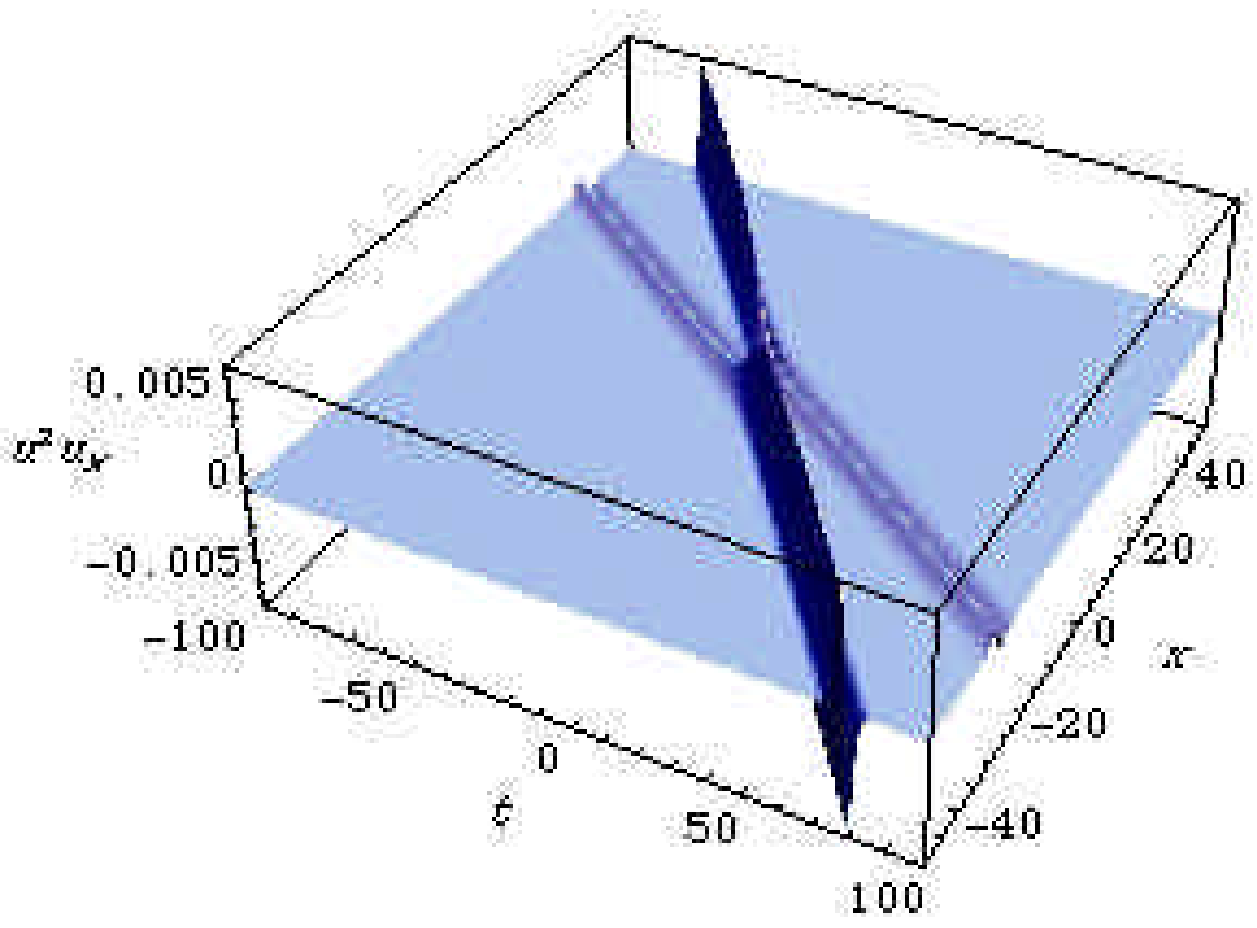

Fig. 2

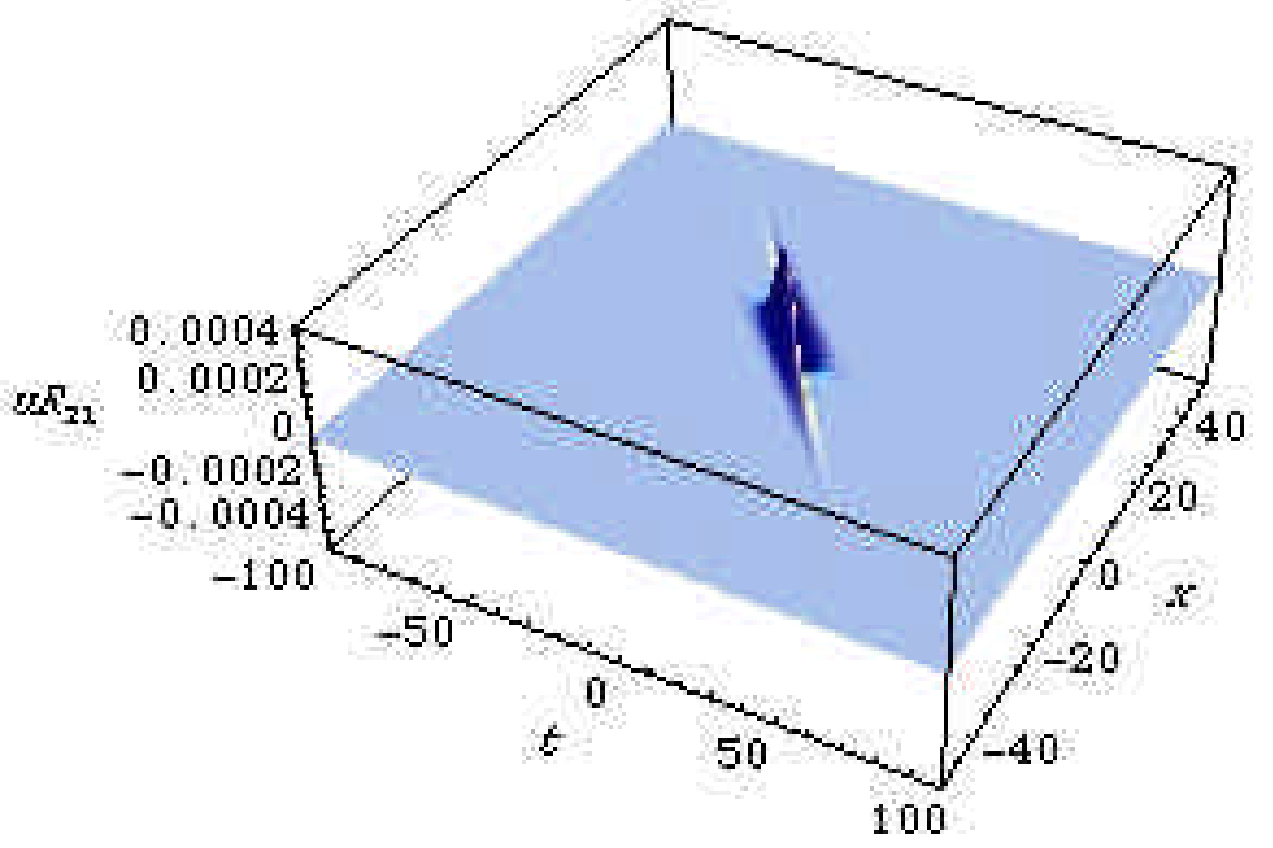

Fig. 3